\documentclass[12pt]{article}
\usepackage[T1]{fontenc}

\makeatletter

\providecommand{\LyX}{L\kern-.1667em\lower.25em\hbox{Y}\kern-.125emX\@}
\let\SF@@footnote\footnote
\def\footnote{\ifx\protect\@typeset@protect
    \expandafter\SF@@footnote
  \else
    \expandafter\SF@gobble@opt
  \fi
}
\expandafter\def\csname SF@gobble@opt \endcsname{\@ifnextchar[%]
  \SF@gobble@twobracket
  \@gobble
}
\edef\SF@gobble@opt{\noexpand\protect
  \expandafter\noexpand\csname SF@gobble@opt \endcsname}
\def\SF@gobble@twobracket[#1]#2{}

\makeatother

\begin{document}

\begin{flushright}  %%For preprint use only
\setlength{\baselineskip}{2.6ex}
TRI--PP--00-03\\
Feb 2000\\
\end{flushright}
\vspace{0.8cm}

\begin{center}
{\huge  Hyperfine effects in charmed baryons }\\
\vspace{0.8cm}
R.M. Woloshyn\\
TRIUMF, 4004 Wesbrook Mall, Vancouver, BC, Canada V6T 2A3
\end{center}
\date{}

\begin{abstract}
Hadron masses are calculated in quenched lattice QCD with a fermion action of 
the D234 type on an anisotropic lattice. Hyperfine splittings for singly 
charmed baryons are found to be in agreement with expectations from the quark
model and with a magnitude slightly larger than experimental values. Masses of 
doubly charmed baryons are also calculated and compared to a variety of model
calculations. Hyperfine splittings in doubly charmed baryons are found
to be slightly smaller than in singly charmed states. 
\end{abstract}
\thispagestyle{empty}\newpage

\section{Introduction}

Relatively little work has been done on heavy baryons using lattice QCD. The
most complete study in a relativistic framework is the one done in the UKQCD
collaboration\cite{ukqcd1}  for both charm and bottom flavoured baryons using an \( O(a) \)
improved fermion action in quenched approximation. In that work
hyperfine splittings in both charm and bottom baryons were found to be considerably
smaller than those predicted by phenomenological models\cite{martin} and
observed experimentally. A recent NRQCD calculation\cite{ali1}
was able to resolve the hyperfine splittings for baryons with b quarks
and found values in accord with phenomenological expectations.
Since the hyperfine splitting is an important feature of the baryon
spectrum further investigation of charmed baryons seems justified.

In this work we present the results of another quenched lattice simulation for
charmed baryons. Due to limitations in computing resources it was not possible
to use the same lattice spacing and volume as used in the UKQCD calculation.
Rather we work on a more coarse lattice (\( \sim  \)0.2fm) with a highly improved
action. Past experience has shown that results of reasonable accuracy may be
obtained with such lattices\cite{coarse}. In order to check the calculation, the spectrum
of baryons in the light quark (u,d,s) sector was calculated at the same time.
As well, meson masses for both heavy and light quarks were calculated. The results
of all these calculations are in reasonable agreement with experimental values
and with the results obtained at small lattice spacing\cite{cppacs,ukqcd2}.

The simulation reported here differs from  \cite{ukqcd1} in two other respects.
In \cite{ukqcd1}
the interpolating operators used for the baryons were taken to have a form which
emphasizes the heavy quark symmetry. Secondly, the spin 1/2 and spin 3/2 \( \Sigma  \)-like
baryons are interpolated by the same operator, a Rarita-Schwinger spin-3/2 field.
As is well known\cite{benmer}, the correlator of such a field propagates both \( J=1/2 \)
and \( J=3/2 \) states and it is the \( J=1/2 \) projection of this correlator
that is identified with the \( \Sigma  \)-like baryons in \cite{ukqcd1}.

The procedure used here is different. As has been done in the context of a QCD
sum rule calculation\cite{bagan}, the interpolating fields used for heavy baryons are taken
to have the same form as those used in the light quark sector. The \( J=1/2 \)
\( \Sigma  \) baryon has an interpolating operator which is distinct from that
used for the \( J=3/2 \) \( \Sigma ^{*} \), just as the nucleon is usually
interpolated by a field that is distinct from the \( \Delta  \). No assumption
is made about heavy quark symmetry. An advantage of this is that we can use
the same procedure (and computer code) at all masses which provides some check
on the results.

The correlation function used to extract the masses of \( \Sigma  \)'s is not
obtained from the projection of the \( J=3/2 \) field's correlator. In fact,
it seems that the spin 1/2 projection of this correlator has very poor overlap
with the \( J=1/2 \) ground state. Our finding is that the spin 1/2 projected
correlator is very small and noisy compared to the correlator calculated directly
using a spin 1/2 field.

The final conclusion of this study is that in quenched lattice QCD
the hyperfine splittings for both
singly and doubly charmed baryons are in reasonable agreement
with phenomenological expectations. Indications are that the splittings may
be overestimated compared to experiment which seems to be a common tendency
for quenched QCD simulations of baryons at all quark masses.

\section{Method}

The calculation is done on an anisotropic lattice\cite{karsch}
using the gauge field action

\begin{eqnarray}
S_{G}(U) & = & \beta \left[ c_{ps}\sum _{ps}(1-\frac{1}{3}Re\, TrU_{ps})+c_{rs}\sum _{rs}(1-\frac{1}{3}Re\, TrU_{rs})\right. \nonumber \\
 &  & +c_{pt}\sum _{pt}(1-\frac{1}{3}Re\, TrU_{pt})+c_{rst}\sum _{rst}(1-\frac{1}{3}Re\, TrU_{rst})\nonumber \\
 &  & \left. +c_{rts}\sum _{rts}(1-\frac{1}{3}Re\, TrU_{rts})\right] 
\end{eqnarray}
 where ps and rs denote spatial plaquettes and spatial planar \( 2\times 1 \)
rectangles respectively. The plaquettes lying in the temporal-spatial planes
are denoted by pt while rectangles with the long side in a spatial(temporal)
direction are labeled by rst(rts). The c coefficients incorporate the aspect
ratio \( \xi =a_{s}/a_{t} \) and gauge link renormalization factors \( u_{s} \)
and \( u_{t} \). These renormalization factors are estimated using the link
expectation value in Landau gauge.

The fermion action is of the anisotropic D234 type\cite{alford} and has the form
\begin{eqnarray}
S_{F} & = & \sum _{x,i}(c_{1i}D_{1i}(x)+c_{2i}D_{2i}(x))+\sum _{x}(c_{1t}D_{1t}(x)+c_{2t}D_{2t}(x))\nonumber \\
 &  & +\sum _{x,i<j}c_{0s}\overline{\psi }(x)\sigma _{ij}F_{ij}(x)\psi (x)+\sum _{x,i}c_{0t}\overline{\psi }(x)\sigma _{0i}F_{0i}(x)\psi (x)\nonumber \\
 &  & -\sum _{x}\overline{\psi }(x)\psi (x),
\end{eqnarray}
where
\begin{equation}
D_{1i}(x)=\overline{\psi }(x)(1-\xi \gamma _{i})U_{i}(x)\psi (x+\widehat{i})+\overline{\psi }(x+\widehat{i})(1+\xi \gamma _{i})U^{\dagger }_{i}(x)\psi (x),
\end{equation}
\begin{equation}
D_{1t}(x)=\overline{\psi }(x)(1-\gamma _{4})U_{4}(x)\psi (x+\widehat{t})+\overline{\psi }(x+\widehat{t})(1+\gamma _{4})U^{\dagger }_{4}(x)\psi (x),
\end{equation}
\begin{eqnarray}
D_{2i}(x) & = & \overline{\psi }(x)(1-\xi \gamma _{i})U_{i}(x)U_{i}(x+\widehat{i})\psi (x+2\widehat{i})\nonumber \\
 &  & +\overline{\psi }(x+2\widehat{i})(1+\xi \gamma _{i})U^{\dagger }_{i}(x+\widehat{i})U^{\dagger }_{i}(x)\psi (x),
\end{eqnarray}
\begin{eqnarray}
D_{2t}(x) & = & \overline{\psi }(x)(1-\gamma _{4})U_{4}(x)U_{4}(x+\widehat{t})\psi (x+2\widehat{t})\nonumber \\
 &  & +\overline{\psi }(x+2\widehat{t})(1+\gamma _{4})U^{\dagger }_{4}(x+\widehat{t})U^{\dagger }_{4}(x)\psi (x).
\end{eqnarray}
 The c coefficients in the fermion action include the aspect ratio, link renormalization
and the hopping parameter factors and are shown in Table \ref{action}.
\begin{table}[top]

\caption{\label{action}Coefficients appearing in the gauge and fermion actions}
{\centering \begin{tabular}{ccccc}
\hline 
&
s&
t&
st&
ts\\
\hline 
\( c_{p\ldots } \)&
\( \frac{5}{3u^{4}_{s}\xi } \)&
\( \frac{5\xi }{3u^{2}_{s}u^{2}_{t}} \)&
&
\\
\( c_{r\ldots } \)&
\( \frac{-1}{12u^{6}_{s}\xi } \)&
&
\( \frac{-\xi }{12u^{4}_{s}u^{2}_{t}} \)&
\( \frac{-\xi }{12u^{2}_{s}u^{4}_{t}} \)\\
\( c_{0\ldots } \)&
\( \frac{2\kappa }{3u^{4}_{s}\xi ^{2}} \)&
\( \frac{2\kappa }{3u^{2}_{s}u^{2}_{t}\xi } \)&
&
\\
\( c_{1\ldots } \)&
 \( \frac{4\kappa }{3u_{s}\xi ^{2}} \)&
\( \frac{4\kappa }{3u_{t}} \)&
&
\\
\( c_{2\ldots } \)&
\( \frac{-\kappa }{6u^{2}_{s}\xi ^{2}} \)&
\( \frac{-\kappa }{6u^{2}_{t}} \)&
&
\\
\hline 
\end{tabular}\par}\end{table}

Hadron masses are calculated from zero-momentum correlation functions in the
usual way. For mesons the interpolating fields were just the standard ones.
For baryons some discussion is needed since the procedure used here differs
from that used in \cite{ukqcd1}. Start from the light quark (u,d,s) sector. A 
common choice\cite{lienweb1}
for the proton operator in terms of u and d quark fields is

\begin{equation}
\label{proton}
\epsilon ^{abc}[u^{T}_{a}C\gamma _{5}d_{b}]u_{c}
\end{equation}
where a,b,c are colour indices and Dirac indices have been suppressed. For
\( \Delta  \) the operator 

\begin{equation}
\label{delta}
\frac{1}{\sqrt{3}}\epsilon ^{abc}\left\{ 2[u^{T}_{a}C\gamma _{\mu }d_{b}]u_{c}+[u^{T}_{a}C\gamma _{\mu }u_{b}]d_{c}\right\} 
\end{equation}
is used. This choice of operators is not unique\cite{ioffe}
but it allows an easy generalization to other baryons\cite{lienweb1}. 
The interpolating operators for the strange hyperons \( \Sigma  \)
and \( \Sigma ^{*} \) are obtained by the replacement \( d\rightarrow s \)
in (\ref{proton}) and (\ref{delta}) respectively. Similarly the interpolators
for \( \Xi  \) and \( \Xi ^{*} \) are constructed by the replacement \( u\rightarrow s \).
The \( \Lambda  \) hyperon is more of a problem. In the SU(3) flavour limit
it would be natural to use the octet lambda
\begin{equation}
\label{octet}
\Lambda _{8}=\frac{1}{\sqrt{6}}\epsilon ^{abc}\left\{ 2[u^{T}_{a}C\gamma _{5}d_{b}]s_{c}+[u^{T}_{a}C\gamma _{5}s_{b}]d_{c}-[d^{T}_{a}C\gamma _{5}s_{b}]u_{c}\right\} 
\end{equation}
 since it is degenerate with the nucleon and \( \Delta  \). However, since
SU(3) flavour is broken, this choice is not compelling. For example, in 
\cite{lienweb1} a combination
\( \Lambda  \) with both SU(3) octet and singlet components was defined.
In this work a ``heavy'' \( \Lambda  \) with a form
\begin{equation}
\label{heavy}
[u^{T}_{a}C\gamma _{5}d_{b}]s_{c},
\end{equation}
 which is natural in the heavy quark limit, is also used.

The operators used to calculate the masses of the eight ground state singly-charmed
baryons are taken to have the same structure as the operators given above. The
\( \Sigma _{c} \) and \( \Sigma ^{*}_{c} \) baryons are obtained by the substitution
\( d\rightarrow c \) in (\ref{proton}) and (\ref{delta}) while \( \Omega _{c} \)
and \( \Omega ^{*}_{c} \) are obtained by \( u\rightarrow s, \) d\( \rightarrow c \).
For the \( \Lambda _{c} \) and \( \Xi _{c} \) the operators (\ref{octet})
and (\ref{heavy}) with the replacements \( s\rightarrow c \) and \( d\rightarrow s, \)
\( s\rightarrow c \) are used. Finally, for the remaining two states we take

\begin{equation}
\label{xiprime}
\Xi '_{_{c}}=\frac{1}{\sqrt{2}}\epsilon ^{abc}\left\{ [u^{T}_{a}C\gamma _{5}c_{b}]s_{c}+[s^{T}_{a}C\gamma _{5}c_{b}]u_{c}\right\} 
\end{equation}
 and
\begin{equation}
\label{xistar}
\Xi _{c}^{*}=\frac{2}{\sqrt{3}}\epsilon ^{abc}\left\{ [u^{T}_{a}C\gamma _{\mu }s_{b}]c_{c}+[s^{T}_{a}C\gamma _{\mu }c_{b}]u_{c}+[c^{T}_{a}C\gamma _{\mu }u_{b}]s_{c}\right\} .
\end{equation}

Rather than using the above operators which have an explicit relativistic form
one could consider the operators which survive in the limit of a static charm
quark. The way to do this has been discussed in some detail in the context of
QCD sum rule calculations\cite{shury,grozin}. 
However there is no particular advantage to
taking this limit here. The operators we use contain the leading heavy-quark
components so the simulation will decide by itself whether they are dominant.
The advantage of using the explicit relativistic forms is that it allows a 
unified analysis of hyperfine effects over the whole mass range from nucleon 
and \( \Delta  \) to charmed baryons.

The operators such as (\ref{delta}) and(\ref{xistar}) propagate both spin
1/2 and spin 3/2 states\cite{benmer}. At zero momentum the correlation function with spatial
Lorentz indices has the general form
\begin{equation}
\label{corfun}
C_{ij}(t)=(\delta _{ij}-\frac{1}{3}\gamma _{i}\gamma _{j})C_{3/2}(t)+\frac{1}{3}\gamma _{i}\gamma _{j}C_{1/2}(t)
\end{equation}
 where the subscripts 3/2 and 1/2 denoted the spin projections. The quantity
\( C_{3/2}(t) \) was used to extract the mass of the spin 3/2 states. However,
it was found that the spin 1/2 projection \( C_{1/2}(t) \) was too noisy at
large time separations to allow for the determination of a mass.

Hadron correlators were calculated using interpolating operators in local form
at both source and sink and also applying a gauge invariant smearing to the
quark propagators at the sink. The Gaussian smearing function, eqn(13) 
of \cite{smear}, was used. Hadron masses were obtained by a simultaneous 
fit to local and sink-smeared correlators.

\section{Results}

The calculations were carried out at \( \beta =2.1 \) on lattices with a bare
aspect ratio \( \xi  \) of 2. The Landau link tadpole factors were first determined
iteratively to be \( u_{s}=0.7858 \) and \( u_{t}=0.9472 \) and these values
were used in all subsequent calculations. The static potential was determined
from both spatial and spatial-temporal Wilson loops. From this the lattice spacing
and renormalized anisotropy were obtained. The results for the lattice spacings
are \( a_{s}^{-1}=(0.977\pm 0.003)GeV \) and \( a_{t}^{-1}=(1.914\pm 0.017)GeV \)
with a systematic uncertainty of \( 0.01GeV \) coming from uncertainty in the
choice of parametrization of the short distance part of the
potential\cite{edwards}. The renormalized
anisotropy was found to be \( 1.95\pm 0.02 \) which is compatible with other
studies done with improved gluon actions at similar lattice 
spacings\cite{renorm}.

Fermion propagators were calculated on a \( 10^{3}\times 30 \) lattice with
Dirichlet boundary conditions on the fermion fields. A total of 420 configurations
were analyzed. With some preliminary tuning it was found that \( \kappa =0.182 \)
and \( \kappa =0.237 \) gave good values for the \( J/\psi  \) and \( \phi  \)
meson masses so these were the \( \kappa  \) values adopted for the charm and
strange quarks for all calculations. Where necessary, masses were extrapolated
to the physical up and down quark region using results from the set of hopping
parameter values \( 0.229,0.233,0.237 \) and \( 0.241. \) The value of the
critical \( \kappa  \) is \( 0.2429(2). \) 

First consider the light quark (u,d,s) sector. The pion and \( \rho  \)-meson
masses were fixed at 0.140GeV and 0.770GeV which determines the hopping parameter
for up and down quarks (taken to be degenerate) and the lattice scale \( a_{\rho } \).
It was found that \( a_{\rho }^{-1}=(1.99\pm 0.12)GeV \) which is slightly
larger than \( a_{t}^{-1} \) found from the static potential. This is an inevitable
result of the quenched approximation. The \( \rho  \)-meson mass scale was
used in all subsequent calculations. The results in the light quark sector are
given in Table \ref{lmasses} with statistical errors obtained by a bootstrap
procedure. The dominant systematic error, a 6\% uncertainty in the scale determination,
is not shown explicitly in this and subsequent tables but should be kept in
mind. For comparison, results from recent calculations (Table II in
\cite{cppacs} and Table XVII in \cite{ukqcd2}) 
done at small lattice spacing and extrapolated to the continuum
are also shown. The results of our coarse lattice simulation are seen to be
quite reasonable. 

\begin{table}[top]

\caption{\label{lmasses}Hadron masses for light quarks. Masses are given in GeV, mass
differences are in MeV. The experimental values in this and other tables are
from \cite{pdg}}
{\centering \begin{tabular}{ccccc}
\hline 
&
This work&
CP-PACS\cite{cppacs}&
UKQCD\cite{ukqcd2}&
Experiment\\
\hline 
\( K \)&
0.485(6)&
0.553(10)&
&
0.498\\
\( K^{*} \)&
0.902(26)&
0.889(3)&
0.748\( ^{+81}_{-46} \)&
0.896\\
\( N \)&
0.942(67)&
0.878(25)&
&
0.940\\
\( \Delta  \)&
1.358(71)&
1.257(35)&
1.25\( ^{+19}_{-9} \)&
1.232\\
\( \Lambda  \)&
1.105(46)&
1.060(13)&
1.088\( ^{+20}_{-19} \)&
1.116\\
\( \Sigma  \)&
1.184(35)&
1.176(11)&
1.091\( ^{+22}_{-11} \)&
1.193\\
\( \Sigma ^{*} \)&
1.488(57)&
1.388(24)&
1.38\( ^{+15}_{-7} \)&
1.384\\
\( \Xi  \)&
1.283(28)&
1.288(8)&
1.242\( _{-24}^{+43} \)&
1.318\\
\( \Xi ^{*} \)&
1.590(39)&
1.517(16)&
1.51\( _{-5}^{+11} \)&
1.534\\
\( K^{*}-K \)&
417(27)&
&
&
398\\
\( \Delta-N  \)&
416(93)&
&
&
292\\
\( \Sigma ^{*}-\Sigma  \)&
304(64)&
&
&
191\\
\( \Xi ^{*}-\Xi  \)&
307(43)&
&
&
216\\
\hline 
\end{tabular}\par}\end{table}

In Table \ref{lmasses} the \( \Lambda  \) mass calculated with the operator
(\ref{octet}) is given. The mass obtained using (\ref{heavy}) was essentially
identical. This was found to be true for all quark masses.

As check on how well charmed quarks are being simulated we first show the results
for charmonium and D-mesons in Table \ref{hmesons}. The hyperfine splitting
between \( \eta _{c} \) and \( J/\psi  \) is \( 78(2)(5)MeV \) which is in
very good agreement with the results obtained in \cite{boyle} using a 
completely different fermion action. 
It is considerably smaller than the experimental value which
is a well known feature of quenched QCD simulations for quarkonium\cite{split}. 
The \( D^{*}-D \)
splittings are compatible with results from NRQCD on similar
lattices\cite{lewis} and are
also smaller than experimental values although the suppression of hyperfine
effects is less pronounced than in charmonium.

\begin{table}[top]

\caption{\label{hmesons}Masses for mesons with charm quarks. Masses are given in GeV,
mass differences are in MeV.}
{\centering \begin{tabular}{ccc}
\hline 
&
This work&
Experiment\\
\hline 
\( \eta _{c} \)&
3.012(4)&
2.980\\
\( J/\psi  \)&
3.087(4)&
3.097\\
\( D \)&
1.875(6)&
1.867\\
\( D^{*} \)&
2.007(9)&
2.008\\
\( D_{s} \)&
1.965(4)&
1.969\\
\( D_{s}^{*} \)&
2.086(6)&
2.112\\
\( J/\psi-\eta _{c}  \)&
78(2)&
117\\
\( D^{*}-D \)&
136(7)&
141\\
\( D_{s}^{*}-D_{s} \)&
122(4)&
143\\
\hline 
\end{tabular}\par}\end{table}

As mentioned in Section 2 the interpolating operators used for charmed baryons
are taken to have the same form as those used for the light baryons. For example,
the correlator for \( \Sigma _{c} \) is calculated directly using operators
which are the same as used for the strange \( \Sigma  \) hyperon except the
mass is increased to charm. An alternative is to extract the masses of \( \Sigma  \)-like
baryons (e.g., \( \Sigma _{c},\Omega _{c} \)) from the spin-1/2 projection
of a correlation function of spin 3/2 fields( see (\ref{corfun})). Figure 1
illustrates why this alternative is not used here. The correlation functions
for \( \Omega _{c} \) and the spin-3/2 projected \( \Omega _{c}^{*} \) are
plotted as a function of lattice time. Also shown is the spin-1/2 projected
correlator. This spin-1/2 projected piece has a very fast pre-asymptotic falloff
and is therefore small and noisy in the time region in which one would want
to extract the mass. This shows that the overlap of the spin-1/2 projection
is small. It is also worth noting from Fig. 1 that even without any analysis
one sees that the \( \Omega _{c} \) correlator has a less rapid falloff than
that of \( \Omega _{c}^{*} \) i.e., \( \Omega _{c} \) is less massive than
\( \Omega _{c}^{*} \).

The results for singly charmed baryons are given in Table \ref{cbmasses}.
Overall the results are in reasonable close to the experimental values where
they are known.

\begin{table}[top]

\caption{\label{cbmasses}Masses of singly charmed baryons. Masses are given in GeV,
mass differences are in MeV. The experimental values are taken from \cite{pdg} 
except for \protect\( \Xi _{c}'\protect \) which is from \cite{jess}}
{\centering \begin{tabular}{cccc}
\hline 
&
This work&
UKQCD\cite{ukqcd1}&
Experiment\\
\hline 
\( \Lambda _{c} \)&
2.304(39)&
2.27 \( ^{+4}_{-3} \) \( _{-3}^{+3} \)&
2.285(1)\\
\( \Sigma _{c} \)&
2.465(22)&
2.46 \( ^{+7}_{-3} \) \( _{-5}^{+5} \)&
2.453(1)\\
\( \Sigma _{c}^{*} \)&
2.557(30)&
2.44 \( ^{+6}_{-4} \) \( _{-5}^{+4} \)&
2.518(2)\\
\( \Xi _{c} \)&
2.454(21)&
2.41 \( ^{+3}_{-3} \) \( _{-4}^{+4} \)&
2.468(2)\\
\( \Xi _{c}' \)&
2.579(14)&
2.51 \( ^{+6}_{-3} \) \( _{-6}^{+6} \)&
2.575(3)\\
\( \Xi _{c}^{*} \)&
2.672(16)&
2.55 \( ^{+5}_{-4} \) \( _{-5}^{+6} \)&
2.645(2)\\
\( \Omega _{c} \)&
2.664(12)&
2.68 \( ^{+5}_{-4} \) \( _{-6}^{+5} \)&
2.704(4)\\
\( \Omega _{c}^{*} \)&
2.757(14)&
2.66 \( ^{+5}_{-3} \) \( _{-7}^{+6} \)&
\\
\( \Sigma _{c}^{*}- \)\( \Sigma _{c} \)&
91(25)&
-17 \( ^{+12}_{-31} \) \( _{-2}^{+3} \)&
65(2)\\
\( \Xi _{c}^{*}-\Xi _{c}' \)&
94(13)&
-20 \( ^{+12}_{-24} \) \( _{-3}^{+2} \)&
70(4)\\
\( \Omega _{c}^{*}-\Omega _{c} \)&
94(10)&
-23 \( ^{+6}_{-14} \) \( _{-2}^{+3} \)&
\\
\hline 
\end{tabular}\par}\end{table}

The masses of doubly charmed baryons were also calculated. There are no experimental
data but a comparison with a selection of model calculations is given in 
Table \ref{ccbmasses} 
\begin{table}[top]

\caption{\label{ccbmasses}Masses of doubly charmed baryons in GeV}
{\centering \begin{tabular}{ccccc}
\hline 
&
\( \Xi _{cc} \)&
\( \Xi _{cc}^{*} \)&
\( \Omega _{cc} \)&
\( \Omega _{cc}^{*} \)\\
\hline 
This work&
3.598(13)&
3.682(20)&
3.697(10)&
3.775(12)\\
Potential model\cite{gersh}&
3.478&
3.61&
&
\\
Mass formulae\cite{burak}&
3.610(7)&
3.735(17)&
3.804(8)&
3.850(25)\\
Bag model\cite{ponce}&
3.511&
3.630&
3.664&
3.764\\
HQET\cite{korner}&
3.610&
3.680&
3.710&
3.760\\
Skyrmion model\cite{rho}&
3.752&
3.793&
3.934&
3.953\\
\hline 
\end{tabular}\par}\end{table}
 (A more complete tabulation from various models may be found in \cite{burak}.) As might
expected, without experimental constraints, the model calculations vary over
a considerable range. The mass splittings from our quenched QCD simulation,
84(13)(5)MeV for \( \Xi _{cc}^{*}-\Xi _{cc} \) and 78(7)(5)MeV for \( \Omega _{cc}^{*}-\Omega _{cc} \),
are substantial and lie in the middle of the range covered by the models listed
in Table \ref{ccbmasses}. They are only slightly smaller than the hyperfine
splittings found for singly charmed baryons.

\section{Summary}

Hadron masses were calculated with an improved action on an anisotropic lattice
with a spatial lattice spacing of about 0.2fm. Comparison with simulations done
at small lattice spacings and extrapolated to the continuum indicate that lattice
spacing errors are less than 10\%.

The focus of this study is charmed baryons.
For both singly
and doubly charmed baryons spin splittings were found to be in agreement
with expectations of quark models and other phenomenological approaches.
The splittings for singly charmed baryons are somewhat larger than experimental
values. This is in contrast to the small
hyperfine effects for singly charmed baryons reported by the 
UKQCD collaboration\cite{ukqcd1}.

The calculations reported here were done on a small lattice at a relatively
large lattice spacing. By doing a unified study for light and heavy quark masses
and comparing to continuum results where possible we have some confidence that
the correct qualitative pattern of hyperfine effects in charmed baryons has
been established for quenched lattice QCD. For a precise calculation, 
finite volume and lattice spacing issues have to be addressed. It is hoped 
that this can be done in the near future.

\section*{Acknowledgements}
It is a pleasure to thank H.R. Fiebig, D.B. Leinweber, R. Lewis, 
N.H. Shakespeare and H.D. Trottier for help and discussion and B.K. Jennings 
for the use of his computers. This work is supported in part by the Natural 
Sciences and Engineering Research Council of Canada.

\newpage
 
%%%%%%%%%%%%%%
%BIBLIOGRAPHY%
%%%%%%%%%%%%%%

\bibliographystyle{unsrt}      %numbering will not stagger if you use 99.

%\newpage

\section*{Figure caption}

\noindent

Fig. 1 Correlation function for the $\Omega_c$ field (squares) and 
 the $\Omega_c^*$ field ( spin 3/2 projection
(triangles), spin 1/2 projection (circles)) as a function of lattice time.

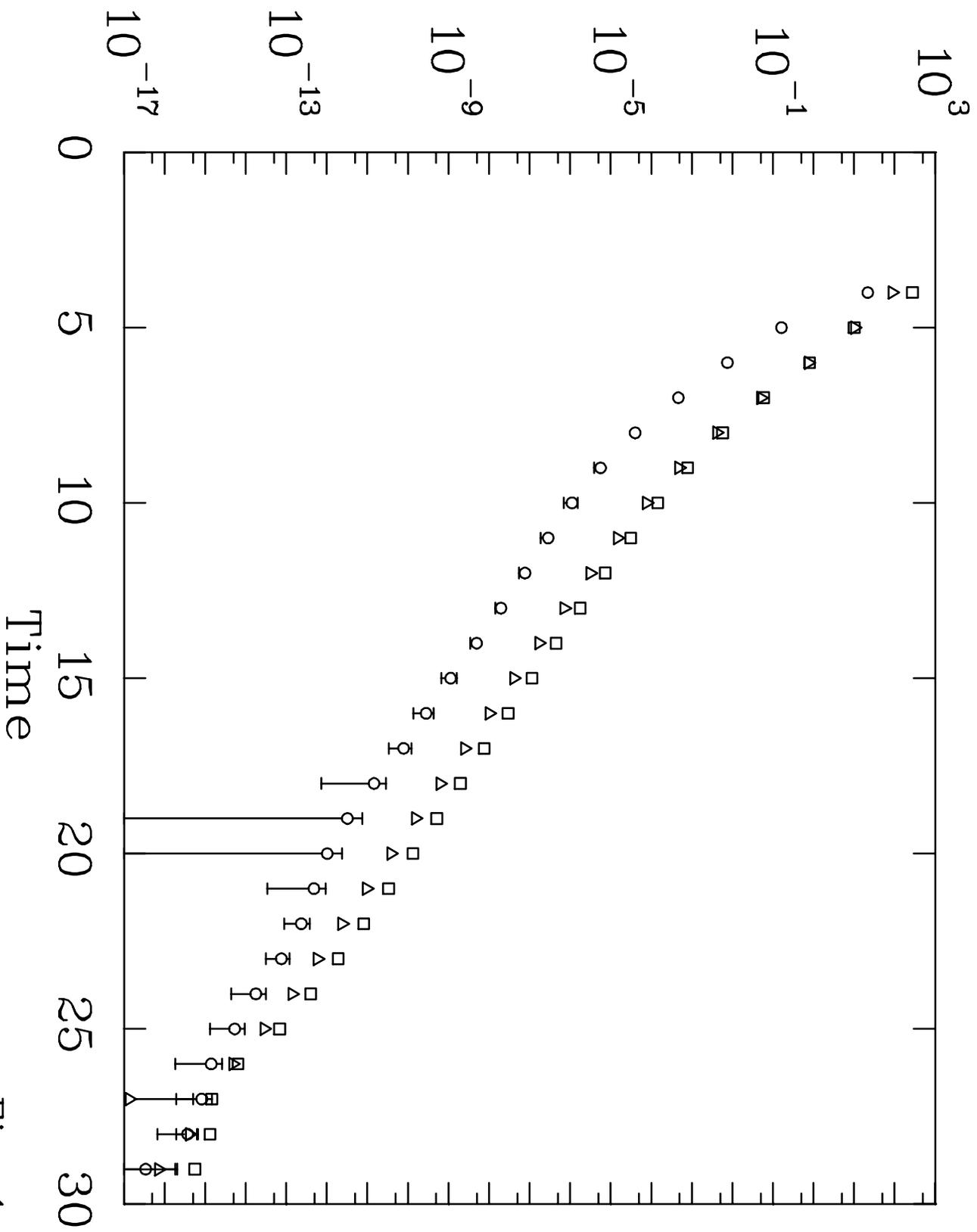
\end{document}